\documentstyle{mn}

\newcommand{\Msun}{\,M$_\odot$}

\def\Msun{\ifmmode{~M_\odot}\else$M_\odot$~\fi}
\def\Teff{\ifmmode{~T_{\mathrm{eff}}}\else$T_{\mathrm{eff}}$~\fi}
\def\kms{\ifmmode{$~km\thinspace s$^{-1}}\else km\thinspace s$^{-1}$\fi}

\def\ga{\mathrel{\mathchoice {\vcenter{\offinterlineskip\halign{\hfil
$\displaystyle##$\hfil\cr>\cr\noalign{\vskip1.5pt}\sim\cr}}}
{\vcenter{\offinterlineskip\halign{\hfil$\textstyle##$\hfil\cr>\cr
\noalign{\vskip1.0pt}\sim\cr}}}
{\vcenter{\offinterlineskip\halign{\hfil$\scriptstyle##$\hfil\cr>\cr
\noalign{\vskip0.5pt}\sim\cr}}}
{\vcenter{\offinterlineskip\halign{\hfil$\scriptscriptstyle##$\hfil
\cr>\cr\noalign{\vskip0.5pt}\sim\cr}}}}}

\def\la{\mathrel{\mathchoice {\vcenter{\offinterlineskip\halign{\hfil
$\displaystyle##$\hfil\cr<\cr\noalign{\vskip1.5pt}\sim\cr}}}
{\vcenter{\offinterlineskip\halign{\hfil$\textstyle##$\hfil\cr<\cr
\noalign{\vskip1.0pt}\sim\cr}}}
{\vcenter{\offinterlineskip\halign{\hfil$\scriptstyle##$\hfil\cr<\cr
\noalign{\vskip0.5pt}\sim\cr}}}
{\vcenter{\offinterlineskip\halign{\hfil$\scriptscriptstyle##$\hfil
\cr<\cr\noalign{\vskip0.5pt}\sim\cr}}}}}

\newif\ifAMStwofonts

\ifoldfss
  \ifCUPmtlplainloaded \else
    \NewTextAlphabet{textbfit} {cmbxti10} {}
    \NewTextAlphabet{textbfss} {cmssbx10} {}
    \NewMathAlphabet{mathbfit} {cmbxti10} {} 
    \NewMathAlphabet{mathbfss} {cmssbx10} {} 
  \fi
  \ifAMStwofonts
    \ifCUPmtlplainloaded \else
      \NewSymbolFont{upmath} {eurm10}
      \NewSymbolFont{AMSa} {msam10}
      \NewMathSymbol{\upi}     {0}{upmath}{19}
      \NewMathSymbol{\umu}     {0}{upmath}{16}
      \NewMathSymbol{\upartial}{0}{upmath}{40}
      \NewMathSymbol{\leqslant}{3}{AMSa}{36}
      \NewMathSymbol{\geqslant}{3}{AMSa}{3E}

       \let\le=\leqslant
       \let\ge=\geqslant
    \fi
  \fi
\fi 

\ifnfssone
  \newmathalphabet{\mathit}
  \addtoversion{normal}{\mathit}{cmr}{m}{it}
  \addtoversion{bold}{\mathit}{cmr}{bx}{it}
  \newmathalphabet{\mathbfit} 
  \addtoversion{normal}{\mathbfit}{cmr}{bx}{it}
  \addtoversion{bold}{\mathbfit}{cmr}{bx}{it}
  \newmathalphabet{\mathbfss} 
  \addtoversion{normal}{\mathbfss}{cmss}{bx}{n}
  \addtoversion{bold}{\mathbfss}{cmss}{bx}{n}
  \ifAMStwofonts
    \ifCUPmtlplainloaded \else
      \UseAMStwoboldmath
      \makeatletter
      \new@mathgroup\upmath@group
      \define@mathgroup\mv@normal\upmath@group{eur}{m}{n}
      \define@mathgroup\mv@bold\upmath@group{eur}{b}{n}
      \edef\UPM{\hexnumber\upmath@group}
      \new@mathgroup\amsa@group
      \define@mathgroup\mv@normal\amsa@group{msa}{m}{n}
      \define@mathgroup\mv@bold\amsa@group{msa}{m}{n}
      \edef\AMSa{\hexnumber\amsa@group}
      \makeatother
      \mathchardef\upi="0\UPM19
      \mathchardef\umu="0\UPM16
      \mathchardef\upartial="0\UPM40
      \mathchardef\leqslant="3\AMSa36
      \mathchardef\geqslant="3\AMSa3E

       \let\le=\leqslant
       \let\ge=\geqslant
    \fi
  \fi
\fi 

\ifnfsstwo
  \DeclareMathAlphabet{\mathbfit}{OT1}{cmr}{bx}{it}
  \SetMathAlphabet\mathbfit{bold}{OT1}{cmr}{bx}{it}
  \DeclareMathAlphabet{\mathbfss}{OT1}{cmss}{bx}{n}
  \SetMathAlphabet\mathbfss{bold}{OT1}{cmss}{bx}{n}
  \ifAMStwofonts
    \ifCUPmtlplainloaded \else
      \DeclareSymbolFont{UPM}{U}{eur}{m}{n}
      \SetSymbolFont{UPM}{bold}{U}{eur}{b}{n}
      \DeclareSymbolFont{AMSa}{U}{msa}{m}{n}
      \DeclareMathSymbol{\upi}{0}{UPM}{"19}
      \DeclareMathSymbol{\umu}{0}{UPM}{"16}
      \DeclareMathSymbol{\upartial}{0}{UPM}{"40}
      \DeclareMathSymbol{\leqslant}{3}{AMSa}{"36}
      \DeclareMathSymbol{\geqslant}{3}{AMSa}{"3E}

       \let\le=\leqslant
       \let\ge=\geqslant
    \fi
  \fi
\fi 

\ifCUPmtlplainloaded \else
  \ifAMStwofonts \else 
    \def\upi{\pi}
    \def\umu{\mu}
    \def\upartial{\partial}
  \fi
\fi

\title{Metallicities and kinematics of G and K dwarfs}

\author[C. Flynn and O. Morell]{Chris Flynn$^{1,3}$ and 
Olof Morell$^{2,4}$\\
$^1$Tuorla Observatory, Piikki\"o, FIN-21500, Finland\\
$^2$Uppsala Observatory, Box 515, Uppsala, S-751 20, Sweden\\
$^3$cflynn@astro.utu.fi\\
$^4$olle.morell@helax.se}
\date{Received ; accepted }

\date{Accepted, Received ; in original form }

\begin{document}

\maketitle

\begin{abstract}

  We have used accurate, spectroscopically determined abundances for G
and K dwarfs of Morell, K\"allander and Butcher (1992) and Morell
(1994), to derive a photometric abundance index for G and K
dwarfs. Broadband Cousins $R-I$ photometry is used to estimate
effective temperature and the Geneva $b_1$ colour to estimate line
blanketing in the blue and hence abundance. Abundances can be derived
in the range $-2.0<$[Fe/H]$<0.5$ for G0 to K3 dwarfs, with a scatter
in [Fe/H] of 0.2 dex.  We apply the method to a sample of Gliese
catalog G and K dwarfs, and examine the metallicity and kinematic
properties of the stars. The stars show the well established
observational features of the disk, thick disk and halo in the solar
neighbourhood.  We find that the distribution of local K dwarf
metallicity is quite similar to local G dwarfs, indicating that there
is a ``K-dwarf'' as well as a G-dwarf problem.

\begin{keywords}
G and K dwarfs -- abundances, kinematics
\end{keywords}

\end{abstract}

\section{Introduction}

  The distribution of metallicity in nearby G-dwarfs, a major
constraint on models of the evolution of the Galaxy, has long
presented us with the ``G dwarf problem'' (Pagel and Patchett
1975). The G-dwarf problem arises because there is an apparent paucity
of metal poor G-dwarfs relative to what we would expect from the
simplest (closed box) models of Galactic chemical evolution. There are
rather a large number of ways that the evolutionary models can be
modified in order to bring them into consistency with the data, such
as pre-enrichment of the gas, a time dependent Initial Mass Function
or gas infall.

  
  G-dwarfs are sufficiently massive that some of them have evolved
away from the main sequence, and these evolutionary corrections must
be taken into account when determining their space densities and
metallicities. While these problems are by no means intractable, it
has long been recognised that K dwarfs would make for a cleaner sample
of the local metal abundance distribution, because for these stars the
evolutionary corrections are negligible.  K dwarfs are of course
intrinsically fainter, and it has not been until recently that
accurate spectroscopic K dwarf abundance analyses have become
available, with which to calibrate a photometric abundance estimator.
Furthermore, with the release of {\it Hipparcos} data expected soon,
accurate parallaxes and distances of a complete and large sample of K
dwarfs will become available, from which the distribution of K dwarf
abundances can be measured. Also, the accurate parallax results given
by {\it Hipparcos} will mean that we can select dwarfs by absolute
magnitude, which is a better way of isolating stars of a given mass
range than is selection by colour (as has been used in samples of G
dwarfs).
 
  In this paper, we have developed a photometric abundance indicator
for G and K dwarfs. In section 1 we have taken a sample of nearby disk
G and K dwarfs for which accurate spectroscopic abundances in a
number of heavy elements and effective temperatures have been measured
by Morell (1994).  We have supplemented these data with several low
metallicity G and K dwarfs for which accurate metallicity and
effective temperature data are available in the literature.  In
sections 2 and 3 we use broadband $VRI$ and Geneva photometry (taken
from the the literature), to develop an abundance index which
correlates well with the spectroscopic metallicities, and can be
transformed to abundance with an accuracy of circa 0.2 dex. In section
4 we measure abundances for approximately 200 G and K dwarfs drawn
from the Gliese catalog.  In sections 5 and 6 we describe the
kinematics of the dwarfs, and demonstrate that the K dwarfs show the
same paucity of metal deficient stars as seen in the G dwarfs,
indicating that there is a ``K dwarf'' as well as a G dwarf
problem. In section 6 we draw our conclusions.

\section{Spectroscopic G and K dwarf Sample}

  Our starting point for calibrating a photometric abundance index for
the G and K dwarfs is a sample of accurate and homogeneously determined
spectroscopic abundances. Good abundances for K dwarfs have been
difficult to carry out until recently, because of the effects of line
crowding, the extended damping wings of strong lines, the strong
effects on the atmospheres of molecular species and the intrinsic
faintness of the stars.

  Our sample of dwarfs comes primarily from Morell, K\"allander and
Butcher (1992) and Morell (1994). These authors give accurate
metallicities, gravities and effective temperatures for 26 G0 to K3
dwarfs. Morell (1994) observed a sample of dK stars with high
dispersion (resolving power 90,000) and high signal to noise with the
ESO Coud\'e Auxiliary Telescope (CAT) at La Silla.  The sample
included all dK stars in the Bright Star Catalogue which were
observable from La Silla, after removing known spectroscopic
binaries. Wavelength regions were primarily chosen to determine CNO
abundances as well as various metals (Na, Mg, Al, Si, Ca, Sc, Ti, V,
Cr, Fe, Co, Ni, Cu, Y, Zr, La, Nd) at 5070.0 to 5108.0 \AA, 6141.0 to
6181.5 \AA, 6290.0 to 6338.0 \AA, 6675.0 to 6724.5 \AA and 7960.0 to
8020.0 \AA. Signal to noise exceeded 100 for most stars and spectral
regions.

  The spectra were analysed using spectral synthesis methods, based on
model atmospheres calculated with the ODF version of the MARCS program
(Gustafsson et. al. 1975). Initial estimates of the stellar effective
temperatures were made from the $V-R$ and $R-I$ colours, using the
temperature scale of vandenBerg and Bell (1985). (Cousins UBVRI
photometry was obtained for 17 stars with the ESO 50 cm telescope in
April and November 1988 and February 1989).  The temperatures were
then improved by examining 12 Fe lines with a range of excitation
energies, and adjusting the temperatures until no trends were seen
between excitation energy and the derived abundance of the species.
For half the stars this lead to adjustments of less than 50 K, and for
the remaining half to adjustments between 50 and 250 K.  Gravities
were determined from a single Ca line at $\lambda\,6162$ \AA. Three of
the G stars in the sample were found to be slightly evolved, with
lower log(g) values. Abundances were determined using spectral
synthesis techniques for many species; here we describe only the Fe
abundances. Fe abundances were measured for 12 neutral, weak and
unblended Fe lines, and very good agreement was obtained amongst the
lines.  The errors in the derived mean Fe abundances are estimated as
smaller than 0.05 dex. An error of approximately 100 K in adopted
effective temperature leads to a change in derived Fe abundance of
only 0.01 dex, so any systematic errors in the temperature scale do
not have a large effect on the abundance scale. Table 1 shows our
sample of G and K dwarfs. Column 1 shows the HD number, column 2 a
secondary identification, column 3 the spectral type Sp, column 4 the
effective temperature \Teff, column 5 the surface gravity log(g), and
column 6 the spectroscopically determined abundance [Fe/H]$_{\rm
Spec}$, with a note on its source in column 7. Columns 8 and 9 show
$b_1$ and Cousins $R-I$, with a note on the source of $R-I$ in column
10. The estimated abundance [Fe/H]$_{\mathrm{Gen}}$ based on $b_1$ and
$R-I$ (as described in the next section) is shown in last column.

\begin{table*}
\small
\caption{The G and K dwarf sample.}
\begin{center}
\begin{tabular}{lllrrrcrrcr}
\hline
  HD      & Other ID& Sp & \Teff   &log(g)& [Fe/H]$_{\rm Spec}$& Note & $b_1~~$ & $R-I$ & Note & [Fe/H]$_{\rm Gen}$    \\
    2151  &HR98   &G2IV  &   5650  &    4.0~~  & $  -0.30~~$& 1&  $  1.072$  & 0.339 &   9&  $-0.33~~~~$    \\
    4628  &HR222  &K2V   &   5150  &    4.6~~  & $  -0.40~~$& 2&  $  1.252$  & 0.443 &   2&  $-0.18~~~~$    \\
   10361  &HR487  &K5V   &   5100  &    4.6~~  & $  -0.28~~$& 2&  $  1.235$  & 0.451 &   2&  $-0.42~~~~$    \\
   10700  &HR509  &G8V   &   5300  &    4.4~~  & $  -0.50~~$& 1&  $  1.129$  & 0.385 &   9&  $-0.45~~~~$    \\
   13445  &HR637  &K1V   &   5350  &    4.6~~  & $  -0.24~~$& 2&  $  1.186$  & 0.420 &   2&  $-0.43~~~~$    \\
   23249  &HR1136 &K0IV  &   4800  &    3.9~~  & $  -0.10~~$& 1&  $  1.263$  & 0.435 &   9&  $ 0.02~~~~$    \\
   26965  &HR1325 &K1V   &   5350  &    4.6~~  & $  -0.30~~$& 2&  $  1.198$  & 0.419 &   2&  $-0.31~~~~$    \\
   38392  &HR1982 &K2V   &   4900  &    4.6~~  & $  -0.05~~$& 2&  $  1.301$  & 0.462 &   2&  $-0.02~~~~$    \\
   63077  &HR3018 &G0V   &   5600  &    4.0~~  & $  -1.00~~$& 1&  $  1.016$  & 0.360 &   9&  $-1.06~~~~$    \\
   72673  &HR3384 &K0V   &   5200  &    4.6~~  & $  -0.35~~$& 2&  $  1.157$  & 0.405 &   2&  $-0.47~~~~$    \\
  100623  &HR4458 &K0V   &   5400  &    4.6~~  & $  -0.26~~$& 2&  $  1.183$  & 0.412 &   2&  $-0.35~~~~$    \\
  102365  &HR4523 &G5V   &   5600  &    4.1~~  & $  -0.30~~$& 1&  $  1.080$  & 0.360 &   9&  $-0.53~~~~$    \\
  131977  &HR5568 &K4V   &   4750  &    4.7~~  & $   0.05~~$& 2&  $  1.420$  & 0.522 &   2&  $ 0.20~~~~$    \\
  136352  &HR5699 &G4V   &   5700  &    4.0~~  & $  -0.40~~$& 1&  $  1.073$  & 0.350 &   9&  $-0.46~~~~$    \\
  136442  &HR5706 &K0V   &   4800  &    3.9~~  & $   0.35~~$& 2&  $  1.372$  & 0.473 &   2&  $ 0.43~~~~$    \\
  146233  &HR6060 &G2V   &   5750  &    4.2~~  & $   0.00~~$& 1&  $  1.092$  & 0.335 &   9&  $-0.11~~~~$    \\
  149661  &HR6171 &K2V   &   5300  &    4.6~~  & $   0.01~~$& 2&  $  1.214$  & 0.397 &   2&  $ 0.10~~~~$    \\
  153226  &HR6301 &K0V   &   5150  &    3.8~~  & $   0.05~~$& 2&  $  1.260$  & 0.450 &   2&  $-0.20~~~~$    \\
  160691  &HR6585 &G3IV/V&   5650  &    4.2~~  & $  -0.10~~$& 1&  $  1.124$  & 0.335 &   9&  $ 0.15~~~~$    \\
  165341  &HR6752A&K0V   &   5300  &    4.5~~  & $  -0.10~~$& 1&  $  1.228$  & 0.455 &   9&  $-0.53~~~~$    \\
  190248  &HR7665 &G7IV  &   5550  &    4.4~~  & $   0.20~~$& 1&  $  1.171$  & 0.345 &   9&  $ 0.41~~~~$    \\
  192310  &HR7722 &K0V   &   5100  &    4.6~~  & $  -0.05~~$& 2&  $  1.255$  & 0.419 &   2&  $ 0.16~~~~$    \\
  208801  &HR8382 &K2V   &   5000  &    4.0~~  & $   0.00~~$& 2&  $  1.300$  & 0.460 &   2&  $ 0.00~~~~$    \\
  209100  &HR8387 &K4/5V &   4700  &    4.6~~  & $  -0.13~~$& 2&  $  1.382$  & 0.510 &   2&  $ 0.04~~~~$    \\
  211998  &HR8515 &F2V:  &   5250  &    3.5~~  & $  -1.40~~$& 1&  $  1.033$  & 0.410   & 9&  $-1.56~~~~$    \\
  216803  &HR8721 &K4V   &   4550  &    4.7~~  & $  -0.20~~$& 2&  $  1.413$  & 0.530   & 2&  $ 0.04~~~~$    \\\hline  
   64090  &BD+31 1684    &sdG2  &   5419  &    4.1~~  & $  -1.7~~~$& 4&  $  1.013$  & 0.41\,~~& 7&  $-1.72~~~~$    \\
  103095  &BD+38 2285    &G8Vp  &   4990  &    4.5~~  & $  -1.4~~~$& 5&  $  1.106$  & 0.437   & 8&  $-1.30~~~~$    \\ 
  132475  &BD$-$21 4009  &G0V   &   5550  &    3.8~~  & $  -1.6~~~$& 7&  $  0.982$  & 0.38\,~~& 7&  $-1.60~~~~$    \\
  134439  &BD$-$15 4042  &K0/1V &   4850  &    4.5~~  & $  -1.57~~$& 6&  $  1.118$  & 0.447   & 9&  $-1.33~~~~$    \\  
  134440  &BD$-$15 4041  &K0V:  &   4754  &    4.1~~  & $  -1.52~~$& 6&  $  1.185$  & 0.478   & 9&  $-1.17~~~~$    \\ 
  184499  &BD+32 3474    &G0V   &   5610  &    4.0~~  & $  -0.8~~~$& 7&  $  1.032$  & 0.36\,~~& 7&  $-0.93~~~~$    \\
  201889  &BD+23 4264    &G1V   &   5580  &    4.5~~  & $  -1.1~~~$& 7&  $  1.031$  & 0.37\,~~& 7&  $-1.06~~~~$    \\
  216777  &BD$-$08 5980  &G6V   &   5540  &    4.0~~  & $  -0.6~~~$& 7&  $  1.079$  & 0.38\,~~& 7&  $-0.80~~~~$    \\
  ---     &BD+29 ~366    &---   &   5560  &    3.8~~  & $  -1.1~~~$& 7&  $  1.010$  & 0.39\,~~& 7&  $-1.49~~~~$    \\
\hline 
\end{tabular}
\end{center}
\begin{flushleft}
~~~~~~~~~~~~~~~1 : Morell, K\"allander and Butcher (1992) \\
~~~~~~~~~~~~~~~2 : Morell (1994) \\
~~~~~~~~~~~~~~~3 : Spite, Spite, Maillard (1984) \\
~~~~~~~~~~~~~~~4 : Gilroy et.al. (1988), Peterson (1980), Rebolo, Molaro and Beckman (1988) \\ 
~~~~~~~~~~~~~~~5 : Sneden and Crocker (1988)\\ 
~~~~~~~~~~~~~~~6 : Petersen (1980), Carney and Petersen (1980) and Petersen (1981) \\ 
~~~~~~~~~~~~~~~7 : Rebolo, Molaro and Beckman (1988) \\
~~~~~~~~~~~~~~~8 : Taylor (1995) \\
~~~~~~~~~~~~~~~9 : Bessell (1990) \\
\end{flushleft}
\end{table*}

  In the next section we develop a photometric abundance index for G
and K dwarfs, which correlates well with the spectroscopic abundances
determined above.  Our aim was to find such an index over as wide a
range of metallicity as possible, so we have supplemented the Morell
data (which are almost all relatively metal rich disk stars) with a
small number of metal weak stars for which spectroscopic metallicities
and effective temperatures have been determined from high dispersion
spectral analyses. These stars were found by searching for metal weak
G and K dwarf stars in the ``Catalog of [Fe/H] Determinations''
(Cayrel de Strobel et. al. 1992), with high dispersion abundance
analyses, and for which Cousins $R-I$ and Geneva photometry could be
located in the literature. The stars are shown in the last 9 rows of
table 1, and come mostly from Rebolo, Molaro and Beckman
(1988). Sources of all the spectroscopic and photometric data are
shown below the table.

\section{Abundance and effective temperature calibration}

  In order to qualitatively understand the effects of [Fe/H] and \Teff
on K dwarfs, a set of synthetic spectra of K dwarf stars over a grid
of [Fe/H] and \Teff was kindly prepared for us by Uffe Gr\aa e J\o
rgensen. As expected, the main effects of metallicity could be seen in
the blue regions (3000 to 4500 \AA) where line blanketing is readily
apparent.

  For all our stars Geneva $(u,b_1,b_2,v_1,v_2,g)$ intermediate band
photometry colours were available in the Geneva Photometric Catalog
(Rufener 1989).  Since Geneva photometry is available for a very large
number of nearby G and K dwarfs, our initial attempt was to develop a
photometric calibration based on Geneva colours only.  However, it
turned out that we could not reliably enough estimate effective
temperature using Geneva photometry, which led to corresponding
uncertainties in the abundance indices we developed. In the end we
used broadband Cousins $RI$ photometry to estimate effective
temperatures, and the Geneva $b_1$ colour to define an index which
measures line blanketing in the blue and correlates well with the
spectroscopic abundances.

  For various plausible colour indices $c_i$ say, being linear
combinations of the six Geneva $(u,b_1,b_2,v_1,v_2,g)$ colours, we
found we could fit linear relations of the form

\begin{equation}
 c_i =  f_i\,{\mathrm{[Fe/H]}_{\mathrm Spec}} + t_i\Teff + a_i.
\end{equation}

  with low scatter (i.e. less than few$\times 0.01$ mag.), where
$f_i, t_i$ and $a_i$ are constants.  For any two such indices, $c_1$
and $c_2$ say, two relations can be inverted to derive a calibration
for [Fe/H] and $T_{\mathrm eff}$. (Note that we also checked for
dependence of each index on log$(g)$, but no significant dependence
was present for any of the indices tried. Hence we only consider \Teff
and [Fe/H] here).  We searched for two indices which were respectively
more sensitive to abundance and to temperature, so the inversion would
be as stable as possible. However, for all the filter combinations we
tried, the linear relations fitted were close to being parallel
planes, which is to say that in the spectral region covered by Geneva
photometry, it is difficult to break the degeneracy between abundance
and effective temperature effects for this type of star.

  Moving to photometry in the near IR was the obvious way around this
problem, since line blanketing is much weaker in this region.  We
gathered $VRI$ photometry from the literature for the stars, and
experimented with the colour indices $V-R$ and $R-I$ ($R$ and $I$ are
throughout the paper on the Cousins system). The $R-I$ data are shown
in the last column of table 1, and are primarily from Morell (1994)
and Bessell (1990).  $R-I$ turned out to have no measurable dependence
on the metal abundance of the stars, and could be used as a very
robust temperature estimator, whereas $V-R$ still showed some
dependency on metallicity. We tried combinations of $R-I$ and Geneva
colours and found that $R-I$ and $b_1$ gave an index which correlated
best with the spectroscopic abundances. (All the Geneva colours were
found to measure line blanketting in the blue and correlated with
metallicity to some extent, with the lowest scatter being for
$b_1$). The relations we fit are:

\begin{equation}
R-I =  1.385 - \Teff/5413.5 
\end{equation}
\begin{equation}
b_1 = 0.121\,{\mathrm{[Fe/H]}_{\mathrm{Spec}}} - \Teff/3480.7 + 2.737.
\end{equation}

  The scatter around the fits $\Delta b_1$ and $\Delta(R-I)$ are shown
as functions of [Fe/H]$_{\mathrm Spec}$, \Teff, log(g), $b_1$ and
$R-I$ in Figure 1. There are no apparent systematic residuals in the
fitting as functions of any of these quantities. In particular, in the
case of $R-I$, there is no dependance on [Fe/H]$_{\mathrm Spec}$ or
log(g), although neither was explicitly fitted, and in the case of
$b_1$, there is no dependence on log(g), although this was not
explicitly fitted.

\begin{figure}
\input epsf
\centering
\leavevmode
\epsfxsize=0.9
\columnwidth
\epsfbox{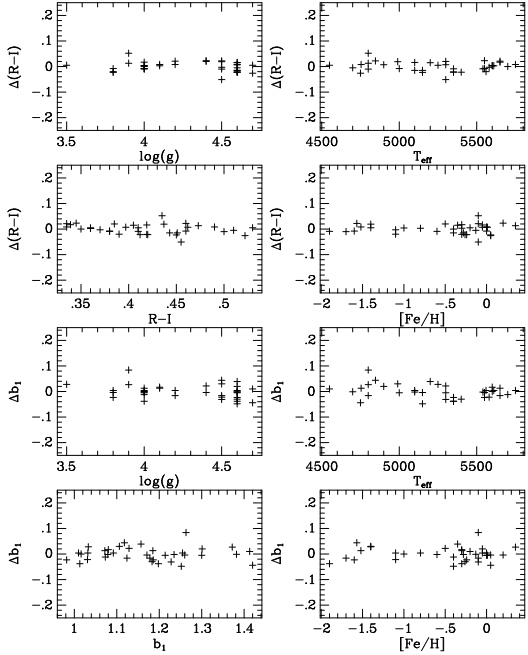}
\caption{The scatter in the fits to $b_1$ and $R-I$ ($\Delta b_1$ and $\Delta
(R-I)$ respectively) is shown as functions of \Teff, [Fe/H], log(g),
$b_1$ and $R-I$. There are no apparent residual trends in the fits in
any of these quantities.}
\end{figure}

Inverting these relations, we derive :

\begin{equation} 
\Teff = 7494. - 5412.\,(R-I) 
\end{equation}
\begin{equation} 
{\mathrm{[Fe/H]}_{\mathrm Gen}} = 8.248\,b_1-12.822\,(R-I)-4.822 
\end{equation}

  Eqns (4) and (5) are valid in the range $0.33 \le R-I \le 0.55$,
which corresponds roughly to G0 to K3 dwarfs.

  Effective temperature calibrations for the $R-I$ filter have been
made by Bessell, Castelli and Plez (1996) from synthetic spectra and
filter band passes, and by Taylor (1992) who used model atmosphere
analyses of Balmer line wings.  We show in Figure 2 Bessell et al.'s
curve (dotted line) for \Teff versus $R-I$ and Taylor's curve (dashed
line), versus our data for the K dwarfs (from table 1).  Our simple
linear fit to the data (Eqn 4) is shown as a solid line. Metal weak
stars ([Fe/H] $<-1.0$ are shown as open squares, showing there is no
systematic difference in temperature scale as a function of
abundance. The match between the data and the three calibrations is
quite satisfactory in the region $0.33 \le R-I \le 0.55$.  For cooler
stars ($R-I \ga 0.55$ i.e. later than about K3) there is a good
indication from the Bessell models that our linear fit cannot simply
be extrapolated outwards. For stars later than about K3 obtaining
accurate abundances from high dispersion spectra becomes increasingly
difficult because of the increasing effects of molecular opacity, and
it was for this reason that the Morell sample stopped at K3. Stellar
atmosphere models and line lists are rapidly improving for cooler
stars however, and it should soon be possible to obtain the
spectroscopic abundances necessary to extend the calibration to cooler
stars still.

\begin{figure}
\input epsf
\centering
\leavevmode
\epsfxsize=0.9
\columnwidth
\epsffile{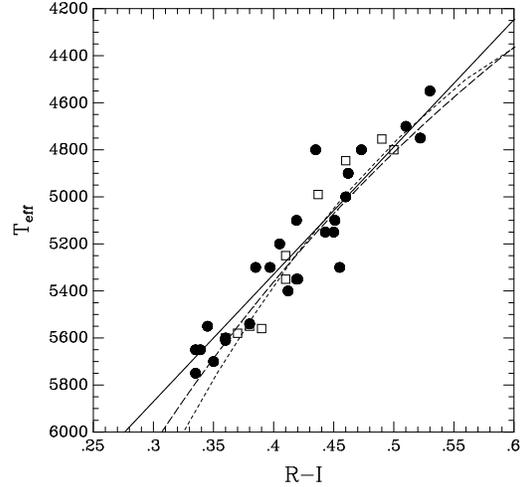}
\caption{\Teff versus $R-I$. The solid line shows our least squares
fit (Eqn. 4), the dotted line the Bessell, Castelli and Plez (1996)
relation based on synthetic spectra, and the dashed line the Taylor
(1992) relation based on analysis of Balmer line wings. Open symbols
are stars with [Fe/H]$<-1.0$.}
\end{figure}

  In figure 3 we show abundances [Fe/H]$_{\mathrm Gen}$ derived using
Eqn. 5 from the $b_1,R-I$ photometry versus the spectroscopically
determined abundances [Fe/H]$_{\mathrm Spec}$ for the stars. The
scatter is 0.18 dex.

\begin{figure}
\input epsf
\centering
\leavevmode
\epsfxsize=0.9
\columnwidth
\epsfbox{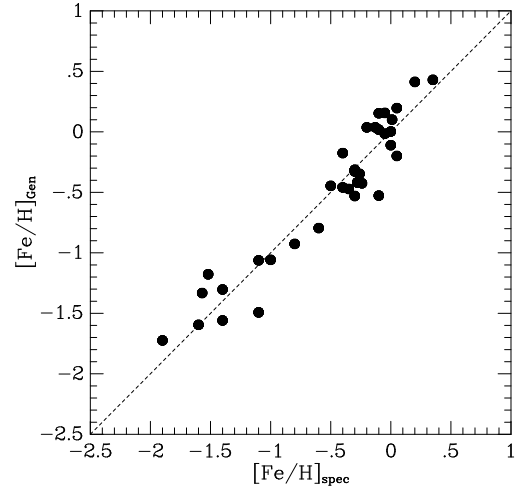}
\caption{Our final abundance calibration, showing the photometric 
abundances [Fe/H]$_{\mathrm Gen}$ versus the spectroscopic abundances
[Fe/H]$_{\mathrm Spec}$ (calculated using Eqn 5).  The line is the 1:1
relation. The scatter for the transformation is 0.18 dex.}
\end{figure}

\begin{figure}
\input epsf
\centering
\leavevmode
\epsfxsize=0.9
\columnwidth
\epsfbox{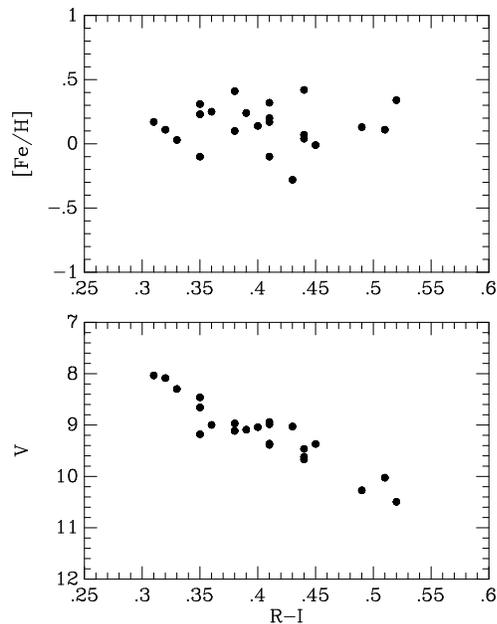}
\caption{The lower panel shows Hyads in the $V$ versus $R-I$ plane. 
The upper panel shows the abundance estimate for each star as a
function of $R-I$ colour. The mean abundance for the stars is [Fe/H]
$=0.14\pm0.03$, with a scatter around the mean of 0.17 dex,
representing the error in an individual measurement.}
\end{figure}

\begin{table}
\small
\caption{Hyades G and K dwarfs.}
\begin{center}
\begin{tabular}{llrrrr}
\hline
  Name   & Sp      & $V~~$ & $R-I$ & $b_1$~~ &  [Fe/H]$_{\rm Gen}$ \\ 
BD +20 598  &  (G5)   & 9.37 &  0.45 &  1.283 &$ -0.01~~~ $ \\       
BD +26 722  &  (G5)   & 9.18 &  0.35 &  1.166 &$  0.31~~~ $ \\       
HD 26756    &  G5V    & 8.46 &  0.35 &  1.117 &$ -0.10~~~ $ \\
HD 26767    &  (G0)   & 8.04 &  0.31 &  1.087 &$  0.17~~~ $ \\       
HD 27771    &  K1V    & 9.09 &  0.39 &  1.220 &$  0.24~~~ $ \\       
HD 28099    &  G8V    & 8.09 &  0.32 &  1.095 &$  0.11~~~ $ \\       
HD 28258    &  K0V    & 9.03 &  0.43 &  1.219 &$ -0.28~~~ $ \\       
HD 28805    &  G8V    & 8.66 &  0.35 &  1.157 &$  0.23~~~ $ \\       
HD 28878    &  K2V    & 9.39 &  0.41 &  1.246 &$  0.20~~~ $ \\       
HD 28977    &  K2V    & 9.67 &  0.44 &  1.274 &$  0.04~~~ $ \\       
HD 29159    &  K1V    & 9.36 &  0.41 &  1.243 &$  0.17~~~ $ \\       
HD 30246    &  (G5)   & 8.30 &  0.33 &  1.101 &$  0.03~~~ $ \\       
HD 30505    &  K0V    & 8.97 &  0.38 &  1.225 &$  0.41~~~ $ \\       
HD 32347    &  (K0)   & 9.00 &  0.36 &  1.174 &$  0.25~~~ $ \\
HD 284253   &  K0V    & 9.11 &  0.38 &  1.188 &$  0.10~~~ $ \\       
HD 284787   &  (G5)   & 9.04 &  0.40 &  1.223 &$  0.14~~~ $ \\       
HD 285252   &  (K2)   & 8.99 &  0.41 &  1.261 &$  0.32~~~ $ \\       
HD 285690   &  K3V    & 9.62 &  0.44 &  1.320 &$  0.42~~~ $ \\       
HD 285742   &  K4V    &10.27 &  0.49 &  1.362 &$  0.13~~~ $ \\       
HD 285773   &  K0V    & 8.94 &  0.41 &  1.210 &$ -0.10~~~ $ \\       
HD 285830   &  ---    & 9.47 &  0.44 &  1.277 &$  0.07~~~ $ \\       
HD 286789   &  ---    &10.50 &  0.52 &  1.434 &$  0.34~~~ $ \\       
HD 286929   &  (K7)   &10.03 &  0.51 &  1.391 &$  0.11~~~ $ \\       
\hline 
\end{tabular}		     	     	       
\end{center}		     	     	       
\end{table}		     	     	          
			     	     	       
\subsection{A check using the Hyades}

  A check of our calibration was made by gathering from the literature
Geneva and $VRI$ photometry for G and K dwarfs in the Hyades
cluster. Cousins $R-I$ colours going well down the main sequence of
the Hyades are available from Reid (1993), and a table of the stars,
their broadband colours and Geneva $b_1$ colour is shown as Table 2.
Figure 4(a) shows the colour magnitude diagram in $V$ versus $R-I$ for
the Hyades G and K dwarfs.  For each star the abundance was estimated
using Eqn 5, and is shown in column 6 in table 2. The abundances are
plotted against the $R-I$ colour in Figure 4(b). The mean abundance of
the stars is [Fe/H] $= 0.14 \pm 0.03$ whit a dispersion of 0.17 dex,
representing the error in an individual measurement.  Taylor (1994)
summarises the literature on the Hyades abundance and gives a best
estimate of [Fe/H]$=0.11\pm0.01$, so our mean abundance of
$0.14\pm0.03$ dex is quite satisfactory. We also note that there is no
indication of a trend of derived Hyades abundances as a function of
$R-I$ colour, so that for these metal rich stars the
temperature-colour relation appears satisfactory.

\section{Abundances and kinematics of Gliese G and K dwarfs}

  The Gliese catalog contains around 800 dwarfs classified between G0
and K3 and estimated to be within 25 pc of the Sun. As a pilot study
for what will be possible after the {\it Hipparcos} data are
available, we have determined abundances for a subset of these stars
having absolute magnitude and space velocity data in the Gliese
catalog and photometric data available in the literature.  The sample
presented here is therefore somewhat inhomogeneous, but the kinematic
and abundance properties of the sample nevertheless allow us to
compare to previous work on G dwarfs using other (abundance estimation
methods) and to show that the G dwarf problem probably extends to the
K dwarfs.

  We obtained the 1991 version of the Gliese catalog (``Preliminary
Version of the Third Catalogue of Nearby Stars'' Gliese and Jahrei\ss)
from the Centre de Donn\'ees astronomiques de Strasbourg and matched
up objects with Bessell's (1990) catalog of $UBVRI$ photometry for
Gliese stars. We selected stars in $0.33 \le R-I \le 0.55$, the colour
range of the abundance calibration. For all these stars $b_1$ data
were obtained from the Geneva catalog.  Some care was needed when
matching the $UBVRI$ and Geneva photometry for stars which were
members of multiple systems to be sure the right components were
matched; all doubtful cases were excluded from further
consideration. Our final lists (184 stars) contained $UBVRI$ and
Geneva photometry, parallax and absolute magnitude data, as well as
$U$, $V$ and $W$ velocities for each star. (Here $U$, $V$ and $W$ are
the usual space motions of the star respectively toward the galactic
center, the direction of galactic rotation and the north galactic
pole.)  Abundances for the stars were derived from the $R-I$ and $b_1$
data using Eqn 5.

\begin{figure}
\input epsf
\centering
\leavevmode
\epsfxsize=0.9
\columnwidth
\epsfbox{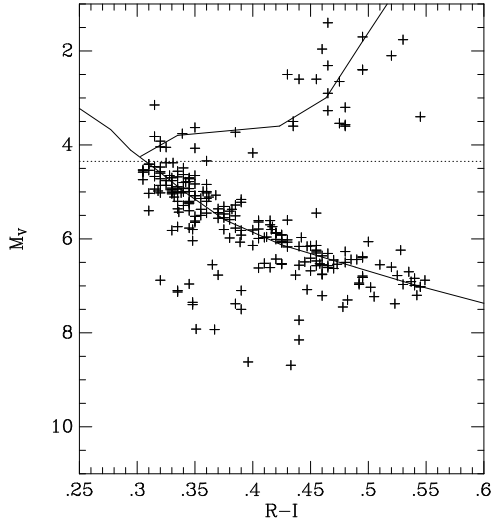}
\caption{Colour magnitude diagram for the Gliese G and K stars. 
The positions of the main sequence and giant branch are shown.  The
dotted line indicates our division into giants (above) and dwarfs
(below) using $M_V=4.35$}
\end{figure}

  Figure 5 shows a colour magnitude diagram for the stars: absolute
visual magnitude $M_V$ versus $R-I$ colour, covering approximately G0
to K3. The solid lines show the positions of the main sequence and the
giant branch in this plane. Giants and dwarfs have been separated
using the dashed line (i.e. $M_V=4.35$ and $0.33 \le R-I \le 0.55$).

  There are a number of stars up to 2 magnitudes below the main
sequence seen in the diagram.  While such stars would classically be
termed subdwarfs, the recent results from the 30 month solution for
{\it Hipparcos} (H30 -- Perryman et.al. 1995) show that the status of
these objects is now uncertain. In fact, most of the objects below the
main sequence are not seen at all in H30 (Perryman et al. their
Figs. 8 (a) and (b)). Perryman et al. ascribe ``a combination of
improved parallaxes, and improved colour indices, the influence of the
latter being particularly important'' as the possible cause.
\footnote{Our measured abundances and the velocities of these ``subdwarfs'' 
indicate that they really are normal disk stars, as H30 appears to
show} In addition, the H30 results indicate that about two thirds of
the stars in the Gliese catalog are not actually within 25 pc, so that
we can expect some changes within Figure 5 after the {\it Hipparcos}
data become available. The data for most of the G and K dwarfs
analysed here are nevertheless of good quality, and we examine their
kinematic and chemical properties next.

\begin{figure}
\input epsf
\centering
\leavevmode
\epsfxsize=0.9
\columnwidth
\epsfbox{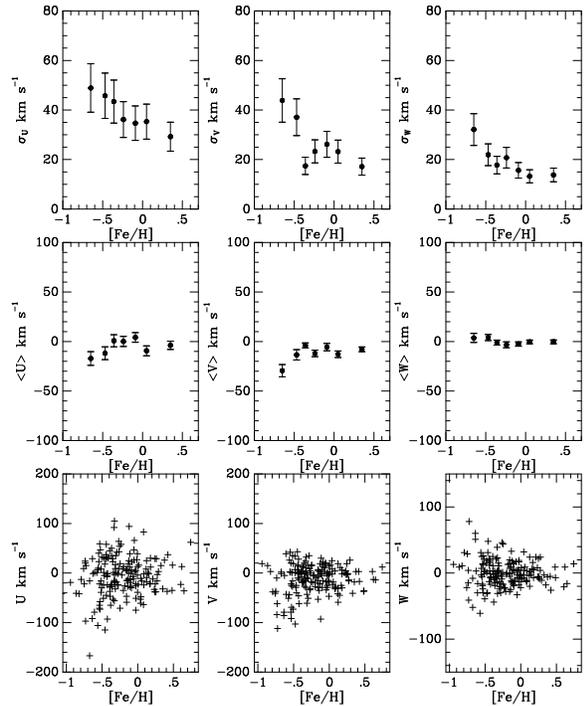}
\caption{Stellar kinematics of Gliese G and K dwarfs as a function of 
abundance. Lower panels show the individual velocities in $U$, $V$ and
$W$ as a function of [Fe/H]. Middle panels show the run of mean
velocity, and the upper panels the run of velocity dispersion with
[Fe/H].}
\end{figure}

  In Figure 6 we show the $U$, $V$ and $W$ velocities of the stars as
a function of [Fe/H], as well as the run of mean velocity $<U>$, $<V>$
and $<W>$ and the velocity dispersions $\sigma_U$, $\sigma_V$,
$\sigma_W$. \footnote{The velocities here have been corrected for a
solar motion of $U_\odot=10$ \kms, $V_\odot=15$ \kms and $W_\odot=8$
\kms (Kerr and Lynden-Bell 1986)} The figure shows the well recognised
properties of the old disk, thick disk and halo as has been seen previously 
in F and G dwarfs and in K giants (see e.g. Freeman 1987, 1996). 

\subsection{Old disk stars}

  The old disk is traced in Figure 6 by stars with
[Fe/H]$\ga-0.5$. The trend in the velocity dispersions is a slow
increase with decreasing abundance, with a possible jump in velocity
dispersion in the thick disk regime $ -1 \la $[Fe/H]$ \la -0.6$. This
behaviour is very similar to that seen in local F and G dwarfs
(e.g. Edvardsson et al 1993) and for K giants (e.g. Janes 1979, Flynn
and Fuchs 1994). For 142 stars with [Fe/H]$ > -0.5$, $(\sigma_U,
\sigma_V, \sigma_W) = (37\pm3, 24\pm2, 17\pm1)$.

\subsection{Thick disk stars}

  Stars in the range $-1 < $[Fe/H]$ \la -0.6$ show a higher velocity
dispersion in all three space motions, and can be identified with the
thick disk. The ratio of thick disk ($-1 < $[Fe/H]$ < -0.6$) to disk
stars ([Fe/H]$ > -0.5$) in the sample is $0.09\pm0.02$, which is the
thick disk local normalisation in this volume limited sample.  This is
in accord with literature estimates of the thick disk local
normalisation, which vary between approximately 2 per cent and 10 per
cent (Reid and Majewski 1993). The elements of the thick disk velocity
ellipsoid\footnote{Note that the star at [Fe/H]$=-0.67, U=-177$ \kms
has been excluded from these calculations, as it may well be a halo
interloper} $(\sigma_U, \sigma_V, \sigma_W) = (45\pm12, 44\pm11,
35\pm9)$ for 16 stars in the range $-1 <$ [Fe/H]$ < -0.6$, and the
assymetric drift is approximately 30 \kms, all in good accord with
previous work (see e.g. Freeman 1987, 1996).

\subsection{Metal weak stars}

  There are 7 stars with [Fe/H] $<-1$, (two of which are bound to each
other, HD 134439 and HD134440 or Gliese 579.2A and B respectively). In
a total sample of 184 stars, 7 halo stars seems rather an
embarrassment of riches. (Bahcall (1986) estimates the local disk to
halo normalisation as 500:1, while Morrison (1993) estimates
850:1). Halo stars are probably over represented in this sample
because they are more likely to be targeted for photometric
observations. It will be interesting to return to the halo and thick
disk normalisations after the {\it Hipparcos} data are available and we can
define and observe a complete volume limited sample of K dwarfs.

\section{The metallicity distribution}

  As discussed in the introduction, the metallicity distribution of
local G dwarfs has long presented a challenge for explaining the
buildup of metallicity in the Galactic disk. We present in this
section the abundance distribution for G dwarfs and for K dwarfs using
our abundance estimator.

  In order to compare to recent work on the G dwarf problem
(e.g. Pagel 1989, Sommer-Larsen 1991, Wyse and Gilmore 1995,
Rocha-Pinto and Maciel 1996), we convert [Fe/H] to [O/H] using the
relations

\begin{flushleft}
\begin{eqnarray}
{\mathrm{[O/H]}} = 0.5{\mathrm{[Fe/H]}} ~~~~~~~~~~ {\mathrm{[Fe/H]}} \ge -1.2\cr
{\mathrm{[O/H]}} = 0.6+{\mathrm{[Fe/H]}} ~~~~~~ {\mathrm{[Fe/H]}} <  -1.2. 
\end{eqnarray}
\end{flushleft}

  When comparing to models, Oxygen abundances are more appropriate
because Oxygen is produced in short lived stars and can be treated
using the convenient approximation of instantaneous recycling -- see
e.g. Pagel (1989).
 
\begin{table}
\small
\caption{Abundance Distributions for G and K dwarfs.}
\begin{center}
\begin{tabular}{rrrrr}
\hline
   [O/H]& $N_G$& $f_G$ & $N_K$ & $f_K$ \\ 
$-1.25$ &   0 & 0.000  &  0 & 0.000  \\
$-1.15$ &   0 & 0.000  &  1 & 0.011  \\
$-1.05$ &   0 & 0.000  &  0 & 0.000  \\
$-0.95$ &   0 & 0.000  &  0 & 0.000  \\
$-0.85$ &   0 & 0.000  &  0 & 0.000  \\
$-0.75$ &   0 & 0.000  &  2 & 0.023  \\
$-0.65$ &   0 & 0.000  &  2 & 0.023  \\
$-0.55$ &   0 & 0.000  &  2 & 0.023  \\
$-0.45$ &   1 & 0.010  &  2 & 0.023  \\
$-0.35$ &   7 & 0.072  &  5 & 0.057  \\
$-0.25$ &  28 & 0.289  & 13 & 0.149  \\
$-0.15$ &  19 & 0.196  & 19 & 0.218  \\
$-0.05$ &  19 & 0.196  & 18 & 0.207  \\
$ 0.05$ &  15 & 0.155  & 10 & 0.115  \\
$ 0.15$ &   6 & 0.062  &  7 & 0.080  \\
$ 0.25$ &   1 & 0.010  &  3 & 0.034  \\
$ 0.35$ &   1 & 0.010  &  3 & 0.034  \\
$ 0.45$ &   0 & 0.000  &  0 & 0.000  \\
\hline
\end{tabular}
\end{center}
The number of G dwarfs, $N_G$ and the number of K dwarfs, $N_K$ binned
by [O/H] in bins 0.1 dex wide, where column 1 indicates the bin
center. There are 97 G dwarfs and 87 K dwarfs. The relative numbers,
normalised to the sample sizes, are shown in the columns headed $f_G$
and $f_K$. See also Figs 7(b) and (c).
\end{table}

  In Figure 7(a) we show the distribution of [O/H] derived using Eqn
6, as a function of $R-I$ colour. The approximate positions of G0, K0
and K3 spectral types are indicated below the stars. Dividing the
stars into G and K dwarfs, we show histograms of the stellar abundance
in the lower panels, normalised to the sample sizes.  In Figure 7(b),
the distribution of [O/H] for 97 G dwarfs shows the well known paucity
of metal poor stars relative to metal rich stars, i.e the G dwarf
problem. The dotted line shows the G dwarf abundance distribution of
Pagel and Patchett (1975) and the dashed line the distribution for
local G dwarfs of Rocha-Pinto and Maciel (1996).  Our histogram
follows the previous determinations well, indicating our abundance
scale is in good agreement with previous work.

  In Fig 7(c) we show the [O/H] distribution for 87 K dwarfs (solid
line). Abundance histogram of this type have been determined by Rana
and Basu (1990) for F, G and K dwarfs, from the 1984 Catalog of
Spectroscopic Abundances (Cayrel de Strobel et al. 1985).  This
procedure suffers from the shortcoming that the abundance data are
inhomogeneous, but we show the abundance distribution of 60 K dwarfs
from Rana and Basu by the dotted line in Fig 7(c). Our distribution
and that of Rana and Basu are in broad agreement, and are similiar to
the abundance distributions for the G dwarfs.

\begin{figure}
\input epsf
\centering
\leavevmode
\epsfxsize=0.9
\columnwidth
\epsfbox{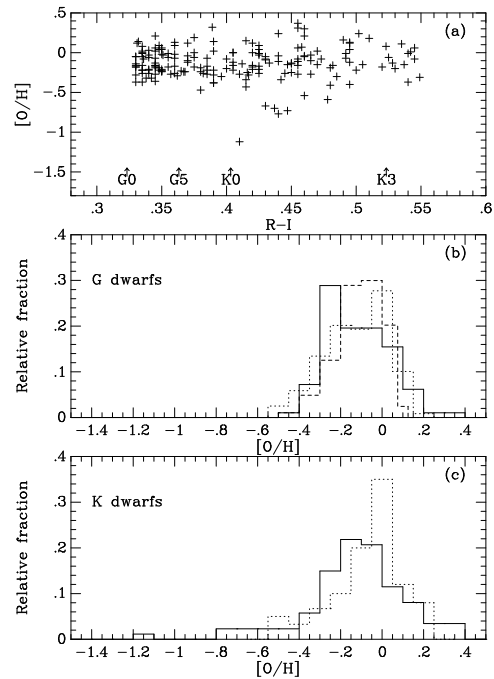}
\caption{Panel (a) Oxygen abundances for the stars as a function of
$R-I$ colour.  Panel (b) shows the histogram of [O/H] in the G dwarfs
in this sample (solid line), for the Pagel and Patchett (1975) sample
(dotted line) and the Rocha-Pinto and Maciel sample (dashed line).
Panel (c) shows our [O/H] distribution for the K dwarfs (solid line)
and that of Rana and Basu (dotted line).}
\end{figure}

  Both the G and K dwarf abundance distributions presented here suffer
from selection bias however.  Since high proper motion or high
velocity stars are frequently targeted for parallax or photometric
observations, metal weak stars are likely to be over-represented, as
was demonstrated for the halo stars ([Fe/H]$<-1$) in section 4.3, and
even stars with thick disk abundances could be
over-represented. Hence, we regard the abundance distributions
reported here cautiously, but nevertheless remark that the K dwarf
abundance distribution is quite similar to that of the G dwarfs, and
offers {\it prima face} evidence for a ``K dwarf problem''. The {\it
Hipparcos} data offer the exciting prospect in the near future of
defining a large, complete and volume limited sample of G and/or K
dwarfs, largely circumventing the above difficulties.  Our abundance
distributions for the G and K dwarfs are shown in Table 3.

  In summary, the kinematics and abundances of the G and K dwarfs
examined in this section follow the trends already well established in
the solar neighbourhood for F and G dwarfs and for K giants. The
ability to measure abundance in K dwarfs offers several interesting
possibilities, especially after the {\it Hipparcos} data become
available. As future work, we plan to analyse the metallicity
distribution of an absolute magnitude selected and volume limited
sample of K dwarfs, which will give us a very clean view of
metallicity evolution in the solar cylinder.

\section{Conclusions}

  We have calibrated an abundance index for G and K dwarfs, which uses
broadband Cousins $R-I$ photometry to estimate stellar effective
temperature, and the Geneva $b_1$ colour to measure line blanketing in
the blue.  Our calibration is based on a recent sample of accurate
abundance determinations for disk G and K dwarfs, (Morell, K\"allander
and Butcher 1992 and Morell 1994) determined from high dispersion
spectra. The index gives [Fe/H] estimates with a scatter of 0.2 dex
for G0 to K3 dwarfs. The [Fe/H] estimator has been checked using the
stars in the Hyades cluster, and we derive a mean abundance for the
stars of [Fe/H]$=0.14$ dex, consistent with previous determinations.
We take a sample of G and K dwarfs from the Gliese catalog, find $R-I$
and $b_1$ data from the literature, and derive the local abundance
distribution for approximately 200 G and K dwarfs. The kinematics of
the G and K dwarfs are examined as function of abundance, the K dwarfs
for the first time, and we see the well known kinematic properties of
the local neighbourhood, as established in the literature from studies
of F and G dwarfs and from K giants. The abundance distributions in
the G and K dwarfs are quite similar, indicating that the ``G dwarf
problem'' extends to the K dwarfs.

\section*{Acknowledgments}
  We thank Bernard Pagel for many helpful discussions and Helio
Rocha-Pinto for interesting comments. This research has made extensive
use of the Simbad database, operated at CDS, Strasbourg, France.
 
{} 


\begin{thebibliography}{} 

\bibitem[]{}Bahcall J.N., 1986, ARA\&A, 24, 577
\bibitem[]{}Cayrel de Strobel G., Bentolila C., Hauck B., Duquennoy A., 1985, A\&AS, 59, 145
\bibitem[]{}Cayrel de Strobel G., Hauck B., Francois P., Thevenin F., Friel E., 
Mermilliod M., Borde S., 1992, A\&AS, 95, 273
\bibitem[]{}Bessell M.S., 1990, A\&AS, 83, 357
\bibitem[]{}Bessell M.S., Castelli F. and Plez B., 1996, preprint 
\bibitem[]{}Carney B. and Petersen R.C., 1980,  ApJ, 244, 989
\bibitem[]{}Flynn C. and Fuchs B., 1994, MNRAS 270, 471 
\bibitem[]{}Freeman K.C., 1987, ARA\&A, 25, 603 
\bibitem[]{}Freeman K.C., 1996, in ``Formation of the Galactic
         Halo... Inside and Out'', Eds. H.~Morrison and A.~Sarajedeni, ASP
         conference Series, Vol. 92, p3.
\bibitem[]{}Gustafsson B., Bell R.A., Eriksson K, Nordlund \AA, 1975, A\&A, 42, 407 
\bibitem[]{}Kerr F. and Lynden-Bell D., 1986, MNRAS, 221, 1023
\bibitem[]{}Gilroy K.K., Sneden C., Pilachowski C.A., Cowan J.J.,
          1988, ApJ, 327, 298
\bibitem[]{}Morell O., 1994, Ph.~D.~Thesis, Uppsala University. Acta Universitatis Upsaliensis
\bibitem[]{}Morell O., K\"allander D. and Butcher H.R., 1992, A\&A, 259, 543
\bibitem[]{}Morrison H., 1993, AJ, 106, 578
\bibitem[]{}Pagel B.E.J., 1989, Rev. Mex. Astr. Astrof. 18, 161
\bibitem[]{}Pagel B.E.J. and Patchett, B.E., 1975, MNRAS, 172, 13
\bibitem[]{}Perryman M.~A.~C., et al., 1995, A\&A,304,69
\bibitem[]{}Petersen R.C., 1980, ApJ, 235, 491
\bibitem[]{}Petersen R.C., 1981, ApJ, 245, 238 
\bibitem[]{}Rana N.C. and Basu S., 1990, Ap\&SS, 168, 317
\bibitem[]{}Rebolo R., Beckman J.E. and Molaro P., 1988, A\&A, 192, 192
\bibitem[]{}Reid I.N., 1993, MNRAS, 265, 785
\bibitem[]{}Reid I.N., Majewski, S.R., 1993, ApJ, 409, 635
\bibitem[]{}Rocha-Pinto H.J. and Maciel W.J., 1996, MNRAS, 279, 447 
\bibitem[]{}Rufener F., 1989, A\&AS, 78, 469
\bibitem[]{}Ryan S., 1992, AJ, 104, 1144
\bibitem[]{}Sneden C. and Crocker D., 1988, ApJ, 335, 906
\bibitem[]{}Sommer-Larsen J., 1991, MNRAS, 249, 368 
\bibitem[]{}Spite M. Spite F. and Maillard, J.P., 1984 A\&A, 141, 56
\bibitem[]{}Taylor B., 1992, PASP 104, 500     
\bibitem[]{}Taylor B., 1995, PASP 107, 734  
\bibitem[]{}vanden Berg D.A. and Bell R.A., 1985, ApJS, 58, 561
\bibitem[]{}Wyse R. and Gilmore G., 1995, AJ, 110, 2771

\end{thebibliography}
\end{document}